\documentclass{article}

\usepackage{PRIMEarxiv}

\usepackage[utf8]{inputenc} % allow utf-8 input
\usepackage[T1]{fontenc}    % use 8-bit T1 fonts
\usepackage{hyperref}       % hyperlinks
\usepackage{url}            % simple URL typesetting
\usepackage{booktabs}       % professional-quality tables
\usepackage{amsfonts}       % blackboard math symbols
\usepackage{nicefrac}       % compact symbols for 1/2, etc.
\usepackage{microtype}      % microtypography
\usepackage{lipsum}		% Can be removed after putting your text content
\usepackage{graphicx}
\usepackage{doi}

\usepackage{amsmath,amsthm,amsfonts,amssymb}
\usepackage{algorithm}
\usepackage{algpseudocode}
\usepackage{setspace}
\usepackage{booktabs}
\usepackage{listings}
\usepackage{color}
\usepackage{subfigure}

\title{Streaming SQL Multi-Way Join Method for Long State Streams
}

\author
{Jinlong Hu *, Tingfeng Qiu \\
  School of Computer Science and Engineering \\
  South China University of Technology, Guangzhou, China\\
  \texttt{jlhu@scut.edu.cn} \\
}

\begin{document}
\maketitle

\begin{abstract}
	%%\lipsum[1]
Streaming computing effectively manages large-scale streaming data in real-time, making it ideal for applications such as real-time recommendations, anomaly detection, and monitoring, all of which require immediate processing. In this context, the multi-way stream join operator is crucial, as it combines multiple data streams into a single operator, providing deeper insights through the integration of information from various sources. However, challenges related to memory limitations can arise when processing long state-based data streams, particularly in the area of streaming SQL. In this paper, we propose a streaming SQL multi-way stream join method that utilizes the LSM-Tree to address this issue. We first introduce a multi-way stream join operator called UMJoin, which employs an LSM-Tree state backend to leverage disk storage, thereby increasing the capacity for storing multi-way stream states beyond what memory can accommodate. Subsequently, we develop a method for converting execution plans, referred to as TSC, specifically for the UMJoin operator. This method identifies binary join tree patterns and generates corresponding multi-way stream join nodes, enabling us to transform execution plans based on binary joins into those that incorporate UMJoin nodes. This transformation facilitates the application of the UMJoin operator in streaming SQL. Experiments with the TPC-DS dataset demonstrate that the UMJoin operator can effectively process long state-based data streams, even with limited memory. Furthermore, tests on execution plan conversion for multi-way stream join queries using the TPC-H benchmark confirm the effectiveness of the TSC method in executing these conversions.
\end{abstract}

% keywords can be removed

\section{Introduction}
With the rapid development of Internet and IoT technologies \cite{88,89,90}, the generation and application of real-time data streams have become increasingly prevalent. Systems such as advertisement analysis \cite{91}, instant delivery services \cite{92}, and financial transactions \cite{93} urgently require the real-time processing and analysis of this data. In light of the dual demands for real-time performance and timeliness, traditional batch processing methods are no longer sufficient. Streaming computing technology, with its ability to process real-time data streams and provide immediate computational responses, plays an increasingly vital role across various fields. Among the diverse range of data stream processing operations, multi-way stream joins are particularly noteworthy for their capacity to address complex queries. However, the process of joining multiple data streams is significantly more intricate and resource-intensive than conventional two-way joins, making in-depth research on multi-way stream joins both practically and academically valuable.

Current popular stream processing frameworks, including Flink and Spark Streaming, implement multi-way stream join processing as binary join trees by default. In these structures, a significant amount of intermediate state generates additional communication and memory overhead, which negatively impacts query efficiency in resource-constrained environments. To address these limitations, existing studies have proposed executing stream joins within a single multi-way join operator, such as the MJoin and MultiStream operators. By eliminating the generation of intermediate states, these operators reduce communication overhead and the need to store intermediate results, demonstrating the potential of multi-way stream join operators as effective solutions for multi-way joins. However, some issues remain to be addressed before multi-way stream join operators can be widely adopted in practice.

One significant issue is that multi-way stream join operators are constrained by memory capacity when processing long state-based data streams. These operators must retain historical input data to perform join calculations, and the substantial volume of input data can lead to memory overflow problems. Current multi-way stream join operators, such as MultiStream, struggle to manage long state-based joins in memory-constrained environments. Flink \cite{9} has implemented long-term state storage based on the Log-Structured Merge-Tree (LSM-Tree) \cite{94} storage structure. Inspired by this, we introduce the LSM-Tree storage structure into the state backend design of multi-way stream join operators and propose a new operator based on LSM-Tree, referred to as UMJoin. This operator utilizes disk storage to manage the long-term state storage of input data streams, enabling the retention of complete historical data beyond memory limits while striving to maintain the integrity of processing results as much as possible.

Additionally, multi-way stream join operators have limitations in their application within streaming SQL, which affects their flexibility. Streaming SQL provides a framework for applying SQL queries and operations to streaming data, enabling users to write queries and transformation operations on real-time data streams using SQL-like syntax. This approach simplifies the development and debugging processes of stream computing, thereby improving overall development efficiency. From the early explorations in systems like Aurora and STREAM to the emergence of stream processing platforms such as Apache Storm and Apache Samza, and the subsequent advancements in modern platforms like Apache Flink, significant progress has been made in this field. With the increasing demand for stream data processing capabilities, supporting complex multi-way stream joins has become a critical focus for the advancement of streaming SQL. In response to this trend, we propose a method for converting execution plans based on streaming SQL, which can transform execution plans derived from binary join trees into plans that incorporate UMJoin nodes, thereby enhancing the flexibility of the UMJoin operator.

In summary, this paper presents the following contributions: (1) We introduce a multi-way stream join operator, UMJoin, specifically designed to process long state-based data streams under memory-constrained conditions. This method facilitates more comprehensive processing of long data streams despite limited memory resources. (2) We propose a method for converting execution plans based on streaming SQL, referred to as the TSC method, which offers a solution for implementing multi-way stream join operators within the context of streaming SQL.

\section{Related Work}
Currently, mainstream stream processing systems, such as Apache Flink \cite{9,44}, Apache Storm \cite{10,45}, and Spark Streaming \cite{11,46}, utilize a multi-way stream join processing strategy based on binary join trees. 

In the early stages of stream processing, the focus was primarily on the simple event processing of a single data stream, including real-time counting, filtering, and basic aggregation. Early stream processing systems, such as Apache Storm, prioritized rapid data stream forwarding and straightforward processing with minimal state management requirements, targeting stateless or ephemeral state applications. As the demand for handling more complex application scenarios grew, stream processing systems began to incorporate state management features. Apache Flink and Spark Streaming exemplify this evolution, offering support for event time and complex aggregation based on time windows. Flink, in particular, introduced "state" as a first-class concept, enabling developers to naturally handle complex state logic across events based on long-term state. Although support for long-term state in stream processing systems has become increasingly sophisticated, current multi-way stream join operators, such as MultiStream, still encounter memory capacity limitations when processing long data streams based on state.

Streaming SQL \cite{47,48,49} is a query language specifically designed for real-time data stream processing. It combines the declarative query syntax of traditional SQL with the capabilities of stream processing technology, allowing users to query and analyze real-time data streams in a manner similar to processing static data. Streaming SQL has continuously evolved and expanded from its early explorations to its current commercial applications. Early stream processing systems, such as Aurora \cite{50} and STREAM \cite{51}, explored efficient data management and query tasks in a streaming environment but did not fully integrate the standard SQL query language. The emergence of stream processing platforms like Apache Storm and Apache Samza \cite{52} enhanced real-time data stream processing capabilities, although they still lacked full support for standard SQL in their initial stages. The introduction of Spark Streaming further advanced stream processing technology, particularly through its Structured Streaming \cite{53,54} extension, which provided improved SQL support for streaming data and enabled developers to perform real-time data processing and analysis in a near-standard SQL manner. Additionally, Apache Flink not only extended support for Streaming SQL but also enhanced the performance and scalability of real-time data processing. The publication of the ISO/IEC 9075:2016 standard \cite{55} marked international recognition of SQL extensions for stream data processing. Recent research by Grandi \cite{48} and Zhang \cite{56} on supporting more complex stream processing operations has further propelled the development of data stream management technology. With the increasing demand for stream data processing capabilities, supporting complex multi-way stream join operators has become a critical focus for the advancement of Streaming SQL.

Although multi-way stream join operators hold significant potential for processing multi-way stream joins, their practical implementation in stream processing systems encounters several challenges. These include limitations in memory capacity when handling long data streams based on state and difficulties associated with using operators based on Streaming SQL. To improve the feasibility and practical significance of implementing multi-way stream join operators in stream processing systems, we have investigated these challenges in depth.

\section{UMJoin in Streaming SQL}
The process of utilizing the UMJoin operator through Streaming SQL is illustrated in Figure 1. When a user submits a Streaming SQL query that involves multi-way joins, the stream processing system parses the SQL text and generates an abstract syntax tree, which encompasses both syntactic and lexical analysis. This abstract syntax tree is then transformed into a logical execution plan that outlines the operations necessary for executing the query. Subsequently, the query optimizer refines the logical execution plan using a cost model, which includes rewriting the query, selecting efficient operators, and choosing the plan with the lowest overall cost to enhance performance. This process results in an initial execution plan structured as a binary tree.

The initial execution plan is converted into a new execution plan that integrates UMJoin nodes using the proposed Two-Step Convert method, namely TSC. This new execution plan is then transformed into a physical operator flow graph, which maps logical operators to physical operators. In this graph, UMJoin nodes are associated with UMJoin physical operators, and the data flow is organized accordingly. The operator flow graph is submitted to physical machines for execution. Data flows through the operators, where it is continuously processed, resulting in the generation of real-time query results. Our primary focus is on the UMJoin operator based on the LSM-Tree and the TSC method for converting Streaming SQL execution plans to incorporate the UMJoin operator.

\begin{figure}[ht]
\label{fig:exec}
  \centering
  \includegraphics[width=0.9\linewidth]{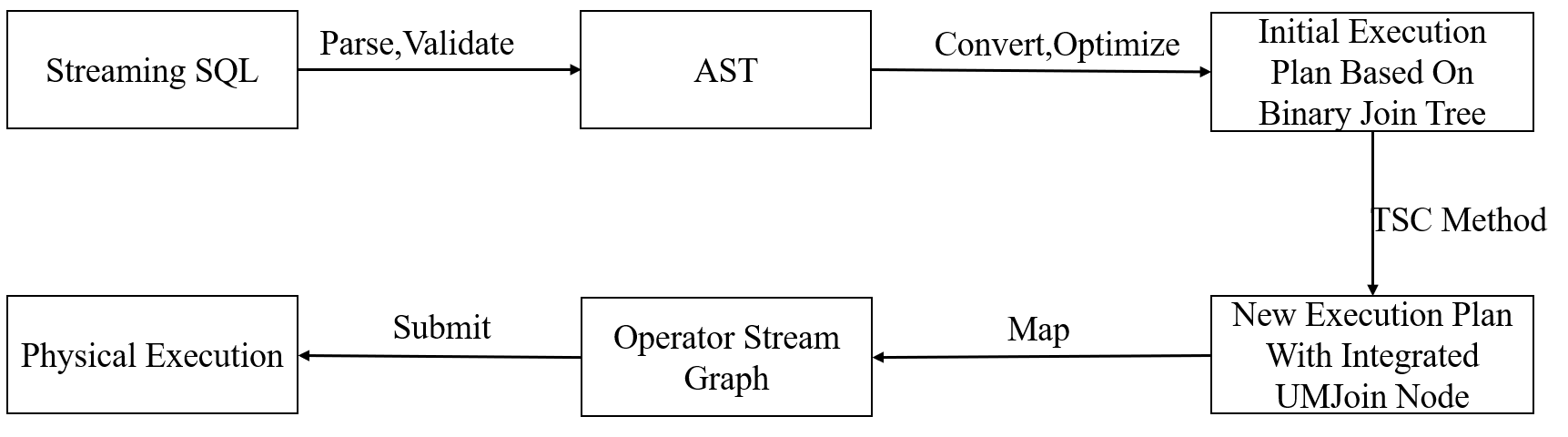}
  \caption{UMJoin in Streaming SQL}  
\end{figure}
\vspace{-15pt}
\subsection{UMJoin}
\subsubsection{Basic Model}
\indent

The UMJoin operator employs a data-driven processing approach that effectively manages join operations involving multiple input data streams. Its basic model is illustrated in Figure 2.

 \begin{figure}[ht]
 \label{fig:umjoin}
  \centering
  \includegraphics[width=0.95\linewidth]{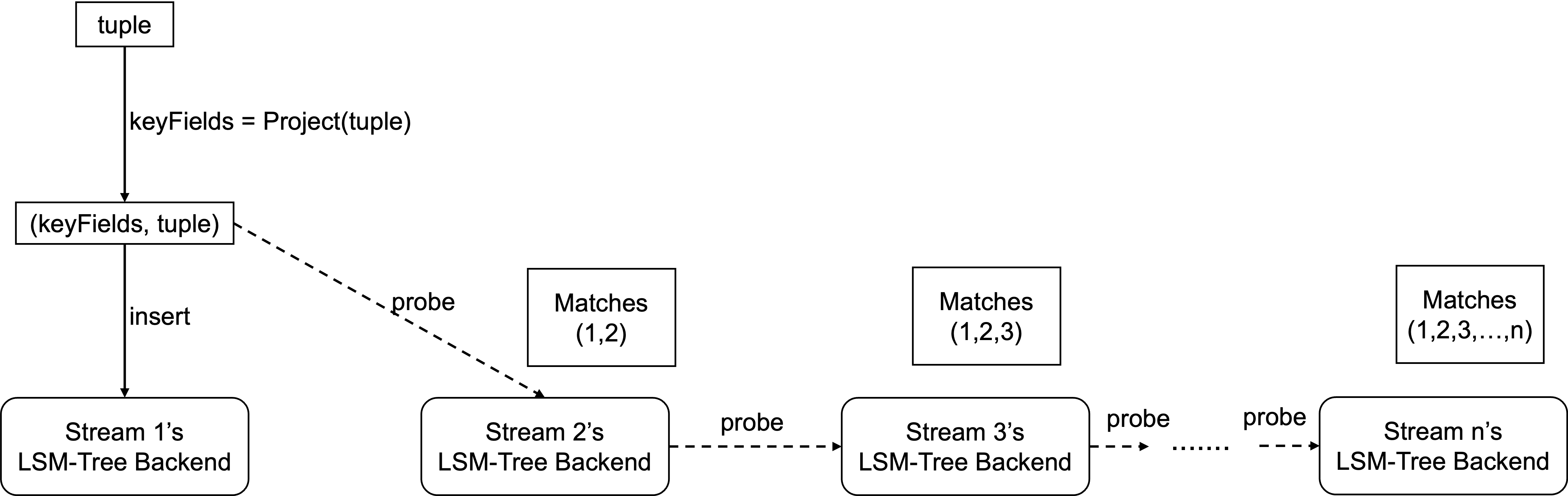}
  \caption{Basic Model of UMJoin}
\end{figure}

For each data stream participating in the join, the UMJoin operator independently establishes a backend utilizing the LSM-Tree storage structure \cite{94,95,96} to facilitate long-term state storage for each data stream input. When a new data tuple arrives at the UMJoin operator, it first performs a projection operation to extract the join key fields from the data tuple. Next, the UMJoin operator executes an insertion operation, placing the data tuple into the corresponding data stream's LSM-Tree backend based on the key fields to ensure storage integrity. Subsequently, the UMJoin operator employs an iterative probing approach, sequentially conducting probing operations on other LSM-Tree backends to identify and concatenate matching data tuples, thereby generating complete join results.

It is important to note that the entire iterative probing process does not necessarily involve all input backends. If a probing operation fails to locate a matching data tuple, the current probing loop is immediately interrupted, and further probing is halted. This design effectively prevents unnecessary computations and resource waste, thereby enhancing overall operational efficiency. In instances where no matching result is found, the UMJoin operator will not produce any output, ensuring both the precision and efficiency of its results.

\subsubsection{LSM-Tree Backend}
\indent

In designing the UMJoin operator to facilitate state-based processing of extensive data streams for multi-way stream joins, we intentionally moved away from the conventional hash table-based in-memory caching strategy to mitigate potential memory overflow issues. Instead, we implemented a key-value pair-based LSM-Tree storage model \cite{94,97} for the long-term persistence of input tuples on disk. This design draws inspiration from the RocksDB state backend utilized in Apache Flink v1.17 \cite{99,98}, extending its LSM-Tree implementation concept to the state backend design of the UMJoin operator. The fundamental structure of the LSM-Tree state backend is illustrated in Figure 3.

\begin{figure}[ht]
  \label{fig:backend}
  \centering
  \includegraphics[width=0.9\linewidth]{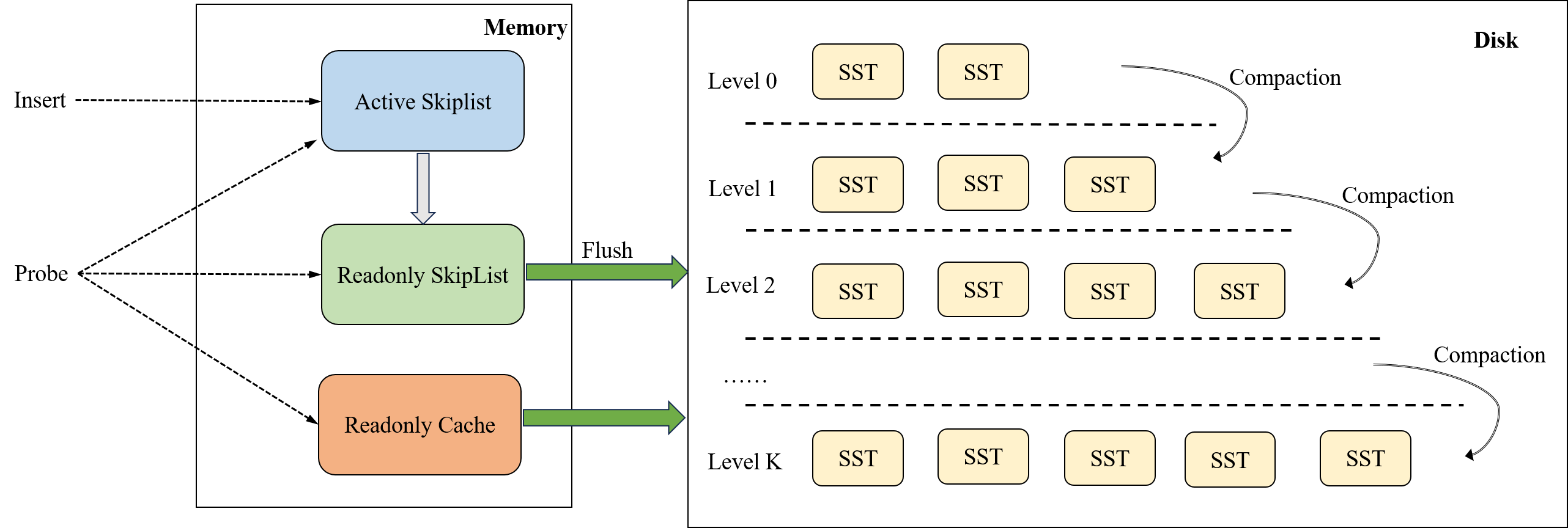}
  \caption{LSM-Tree Backend}
\end{figure}

The LSM-Tree backend consists of both memory and disk layers. The memory layer utilizes a skip list to implement immediate data caching, where each entry in the skip list comprises a unique join key and an associated list of tuples. The advantage of skip lists lies in their ability to support efficient data insertion and sequential access, making them ideal for real-time data processing. The skip list employed in the memory layer is referred to as the active skip list. As the volume of data increases and reaches a predefined threshold, a flush process is initiated, transferring the data in an orderly manner to disk layer files. To minimize the impact on real-time write performance during this data spill process, the active skip list is converted to a read-only state when preparing to flush to disk, and a new active skip list is immediately created to continue accepting new data writes. This separation of read and write operations ensures sustained high throughput in data processing.

Additionally, the memory layer includes a read-only block cache designed to store data blocks that have already been read from the disk, managed using a least recently used (LRU) strategy. When a data block is accessed, it is loaded into this cache. Subsequent read operations that require the same data block can directly retrieve it from the memory cache, significantly enhancing read performance. While increasing the size of the read-only block cache can further improve read performance, it also consumes additional memory resources. Therefore, maintaining an appropriately sized read-only block cache is essential for optimizing the performance of the LSM-Tree backend.

On the disk storage layer, the LSM-Tree backend utilizes a tiered storage mechanism, organized into multiple levels ranging from Level 0 to Level K. Each level consists of several Sorted String Table (SST) files\cite{97}, with the data volume at each level being ten times greater than that of the preceding level. Each SST file contains multiple sorted data blocks and an index block for binary search, with each block sorted by key values. Level 0 SST files are directly generated from the in-memory read-only skip list through a process known as "flush." When the number of Level 0 SST files reaches a specific threshold, they are merged with Level 1 files to create new Level 1 SST files through a process called "compaction." Similarly, higher-level SST files are produced by merging lower-level files with existing files at the current level. During the SST file compaction process, all write operations are executed sequentially to ensure write efficiency and maintain data order. Once an SST file is created, it becomes a read-only data structure and remains unmodified throughout its lifecycle, being selected for deletion during subsequent merges to accommodate newly merged SST files. This design guarantees the stability and durability of data at the disk layer while simplifying data management and merge operations.

When the UMJoin operator processes data streams, insert operations consistently target the active skip list, ensuring immediacy and continuity. For probing operations, the operator queries data based on its heat level, beginning with the active skip list in memory, then proceeding to the read-only skip list, and finally to the read-only block cache, before progressively exploring each level of sorted string tables (SSTs) on the disk. Reading SST files involves locating the appropriate file, utilizing the index and Bloom filter to quickly identify data blocks, and then retrieving the corresponding data blocks from the disk. The read data blocks may be added to the read-only block cache for faster access during subsequent reads. This systematic, hierarchical query strategy not only enhances data retrieval efficiency but also ensures that the UMJoin operator can swiftly respond to multi-way stream join queries.

As data continues to flow in, the disk layer state storage tables (SSTs) accumulate information. To manage the volume of long-term stored data and prevent state overflow, state items that exceed a predetermined time threshold are evaluated and confirmed based on business rules regarding retention necessity. This hybrid expiration strategy ensures that even in long-running environments, the UMJoin operator can effectively manage its state storage and optimize the utilization of storage space.

The LSM-Tree backend design of the UMJoin operator addresses the memory capacity limitations of state storage, facilitating state-based management of extensive data streams. This design enables the UMJoin operator to efficiently process and maintain large-scale dynamic data streams in memory-constrained environments, offering a robust and scalable solution for real-time multi-way stream joins while ensuring the completeness of the processing results from the UMJoin operator.

\subsubsection{Integrity Proof}
\indent

Assuming that the physical disk has sufficient capacity and that the backend storage can accommodate the full volume of input data streams, while disregarding data expiration, this subsection theoretically demonstrates the integrity of the computed results from the UMJoin operator. Consider $n$ input data streams ${S_1}, {S_2}, {S_3}, ..., {S_n}$, with their multi-way join result defined as $R$. This result is obtained through the multi-way join operation, as shown in Equation \ref{eq:c1}, where $ \times $ denotes the join operation.

\begin{equation}
R = {S_1} \times {S_2} \times {S_3} \times ... \times {S_n}
\label{eq:c1}
\end{equation}

Consider that these data streams are incrementally input in an arbitrary order ${o_1}, {o_2}, {o_3}, ..., {o_n}$ over a certain period, with the incremental inputs denoted as $d{S_{{o_i}}}$, assuming they do not overlap in time. When the data stream ${S_{{o_i}}}$ receives new data $d{S_{oi}}$, the cumulative amount of input for each data stream $S_{{o_i}}^*$ is the sum of the original input ${S_{{o_i}}}$ and the incremental input $d{S_{{o_i}}}$, as shown in Equation \ref{eq:c2}.

\begin{equation}
S_{{o_i}}^* = {S_{{o_i}}} + d{S_{{o_i}}}
\label{eq:c2}
\end{equation}

The actual output result increment of the UMJoin operator is $d{R_{{o_i}}}$, as shown in Equation \ref{eq:c3}.

\begin{equation}
d{R_{{o_i}}} = S_{{o_1}}^* \times ... \times S_{{o_{i - 1}}}^* \times d{S_{{o_i}}} \times {S_{{o_{i + 1}}}}... \times {S_{{o_n}}}
\label{eq:c3}
\end{equation}

The theoretical output result ${R^*}$ after all incremental inputs is calculated as shown in Equation \ref{eq:c4}.

\begin{equation}
\begin{array}{l}
{R^*} = S_{{o_1}}^* \times S_{{o_2}}^* \times S_{{o_3}}^* \times  \ldots  \times S_{{o_n}}^*\\
 = ({S_{{o_1}}} + d{S_{{o_1}}}) \times ({S_{{o_2}}} + d{S_{{o_2}}}) \times ({S_{{o_3}}} + d{S_{{o_3}}}) \times  \ldots  \times ({S_{{o_n}}} + d{S_{{o_n}}})\\
 = {S_{{o_1}}} \times {S_{{o_2}}} \times {S_{{o_3}}} \times  \ldots  \times {S_{{o_n}}} + d{S_{{o_1}}} \times ({S_{{o_2}}} + d{S_{{o_2}}}) \times ({S_{{o_3}}} + d{S_{{o_3}}}) \times  \ldots  \times ({S_{{o_n}}} + d{S_{{o_n}}})\\
 + {S_{{o_1}}} \times d{S_{{o_2}}} \times ({S_{{o_3}}} + d{S_{{o_3}}}) \times  \ldots  \times ({S_{{o_n}}} + d{S_{{o_n}}}) +  \ldots  + {S_{{o_1}}} \times {S_{{o_2}}} \times {S_{{o_3}}} \times  \ldots  \times d{S_{{o_n}}}
\end{array}
\label{eq:c4}
\end{equation}

Based on ${R^*} = R + dR$, the theoretical result increment $dR$ can be derived as shown in Equation \ref{eq:c5}.

\begin{equation}
\begin{array}{l}
dR = d{S_{{o_1}}} \times ({S_{{o_2}}} + d{S_{{o_2}}}) \times ({S_{{o_3}}} + d{S_{{o_3}}}) \times  \ldots  \times ({S_{{o_n}}} + d{S_{{o_n}}})\\
 + {S_{{o_1}}} \times d{S_{{o_2}}} \times ({S_{{o_3}}} + d{S_{{o_3}}}) \times\ldots  \times ({S_{{o_n}}} + d{S_{{o_n}}}) + {S_{{o_1}}} \times {S_{{o_2}}} \times d{S_{{o_3}}} \times  \ldots  \times ({S_{{o_n}}} + d{S_{{o_n}}})\\
 +  \ldots  + {S_{{o_1}}} \times {S_{{o_2}}} \times {S_{{o_3}}} \times  \ldots  \times d{S_{{o_n}}}
\end{array}
\label{eq:c5}
\end{equation}

By observing, it can be seen that each term on the right-hand side of the equation corresponds to a $d{R_{{o_i}}}$. Therefore, $dR$ can be expressed as the sum of all individual input increments $d{R_{{o_i}}}$, as shown in Equation \ref{eq:c6}.

\begin{equation}
dR = \sum\limits_{i = 1}^n d{R_{{o_i}}}
\label{eq:c6}
\end{equation}

This demonstrates that the cumulative sum of the result increments calculated by the UMJoin operator at each individual input time is equal to the theoretical result increment over the entire period. This indicates that the UMJoin operator preserves the integrity of its computed results, as long as the backend storage can accommodate the full volume of data.

\subsection{Execution Plan Conversion}
To enable the UMJoin operator in stream SQL, a method called Two-Step Convert (TSC) is proposed to transform execution plans based on binary join trees into plans that incorporate UMJoin nodes. The TSC algorithm, illustrated in Algorithm 1, is executed in two primary steps. The first step involves invoking the \texttt{createMultiJoinGroups()} method to identify patterns of binary join trees within the execution plan and encapsulate them into corresponding multi-way stream join node groups. The second step entails calling the \texttt{getMultiJoinNode()} method, which leverages the pattern recognition results from the first step to transform the identified multi-way stream join groups in the original execution plan into corresponding multi-way stream join nodes, thereby replacing them in the execution plan. This process results in a new execution plan that employs multi-way stream join nodes, specifically UMJoin nodes. The algorithm defines a `MultiJoinGroup` class, which encapsulates a set of nodes that collectively form a valid binary join tree pattern. This class includes a member node list and a reference to a core root node. The constructor \texttt{MultiJoinGroup(root)} sets the provided node as the root node and initializes the member node list for subsequent additions. Figure 4 illustrates a concrete example of the TSC method.

\begin{algorithm}[H]
\caption{Two-Step-Convert}
\label{a1}
\begin{algorithmic}[1]
\Require $rootNode$
\Ensure  $newRootNode$
\State // Step 1: Identify binary join tree patterns and create multi-way join node groups
\State ${\rm{createMultiJoinGroups}}(rootNode)$     
\State // Step 2: Create multi-way join nodes and replace corresponding multi-way join node groups
\State $newRootNode \gets {\rm{getMultiJoinNode}}(rootNode)$  
\State \textbf{return} $newRootNode$
\end{algorithmic}
\end{algorithm}

\begin{figure}[ht]
  \centering
  \includegraphics[width=0.9\linewidth]{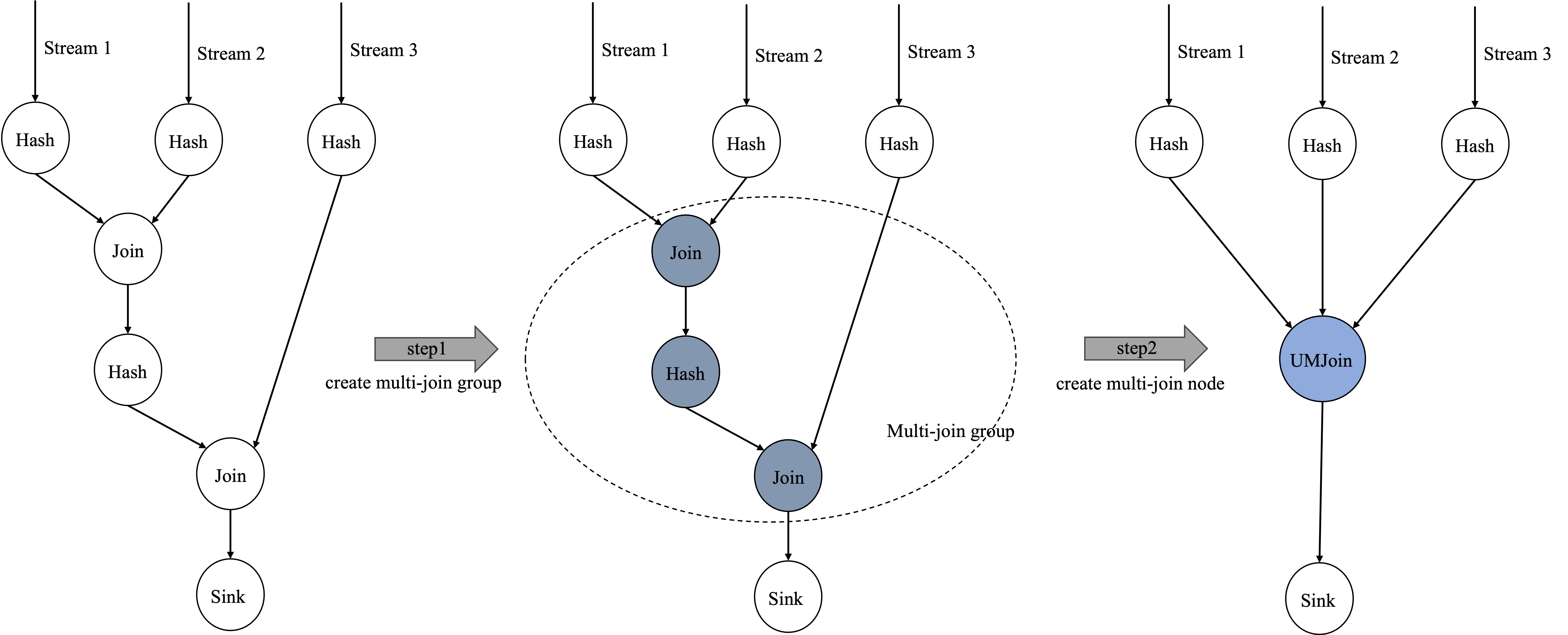}
  \caption{TSC example}
  \label{fig:tsc}
\end{figure}
\vspace{-15pt}

\subsubsection{Creation of Multi-Way Stream Join Node Groups}
The first step of the TSC method involves identifying binary join tree patterns and creating multi-way stream join node groups. As outlined in Algorithm 2, the algorithm takes the root node, denoted as $rootNode$, of the execution plan as input. It begins by performing a breadth-first traversal of $rootNode$ using the \texttt{getOrderedNodes()} method to obtain an ordered list of nodes, referred to as $orderedNodes$ (line 2). The algorithm then iterates over each node in the list, executing a series of operations on each node (lines 3-16). It employs the \texttt{canBeMultiJoinGroupMember()} method (lines 34-35) to determine whether a node satisfies the criteria to become a member of a multi-way join node group (line 4). This method identifies and establishes the binary join tree pattern based on the type of node and its input nodes. An example implementation of the \texttt{canBeMultiJoinGroupMember()} method is provided: if the node is a \texttt{JoinNode} or a \texttt{HashNode} and its first input is a \texttt{JoinNode}, then the node meets the criteria. Nodes that do not meet the criteria are skipped, allowing the algorithm to continue with the next node (line 5). For nodes that meet the criteria, the algorithm further invokes the \texttt{canBeInSameGroupWithOutputs()} function to determine whether the node can be added to an existing multi-way join group (line 7). This decision is based on the consistency of the node’s output connections: if all outputs of the node belong to the same multi-way stream join group, then the node can also be added to that group (lines 18-32). If the \texttt{canBeInSameGroupWithOutputs()} function returns a non-empty multi-way stream node group, the current node is added to the group's member list (lines 8-10). It is important to note that the \texttt{canBeMultiJoinGroupMember()} function is crucial for pattern recognition. The example provided corresponds to the binary join tree pattern for the UMJoin operator. This function can be modified to support additional binary join tree patterns, provided that these patterns correspond to actual multi-way stream join physical operators.

\begin{algorithm}
\caption{createMultiJoinGroups}
\label{a2}
\begin{algorithmic}[1]
\Require $rootNode$
\Function{createMultiJoinGroups}{$rootNode$}
    \State $orderedNodes \gets \text{getOrderedNodes}(rootNode)$ 
    \ForAll{$node$ in $orderedNodes$}
        \If{\textbf{not} \Call{canBeMultiJoinGroupMember}{$node$}}
            \State \textbf{continue}
        \EndIf
        \State $outputGroup \gets \text{canBeInSameGroupWithOutputs}(node)$
        \If{$outputGroup$ is not null}
            \State \textbf{add} $node$ \textbf{to} $outputGroup.\text{members}$
            \State \textbf{continue}
        \EndIf
        \If{$node$ is JoinNode}
            \State $group \gets \text{new MultiJoinGroup}(node)$
            \State $node.\text{group} \gets group$
        \EndIf
    \EndFor
\EndFunction
\Statex
\Function{canBeInSameGroupWithOutputs}{$node$}
    \If{$node.\text{outputs}$ is empty}
        \State \Return null
    \EndIf
    \State $outputGroup \gets node.\text{outputs}[0].\text{group}$
    \If{$outputGroup$ is null}
        \State \Return null
    \EndIf
    \ForAll{$outputNode$ \textbf{in} $node.\text{outputs}$}
        \If{$outputNode.\text{group}$ is not $outputGroup$}
            \State \Return null
        \EndIf
    \EndFor
    \State \Return $outputGroup$
\EndFunction
\Statex
\Function{canBeMultiJoinGroupMember}{$node$}
    \State \Return $node$ is JoinNode or ($node$ is HashNode and $node.\text{inputs}[0]$ is JoinNode)
\EndFunction
\end{algorithmic}
\end{algorithm}

\subsubsection{Creation of Multi-Way Stream Join Nodes}
\indent
In the second step of the TSC algorithm, the identified multi-way stream node groups are converted into multi-way stream join nodes, which are subsequently integrated into the execution plan. Algorithm 3 outlines this process, where the input is the node to be processed, denoted as \(node\), and the output is the transformed node, referred to as \(ret\). The transformation is achieved by recursively traversing each node and its input nodes. In the \texttt{getMultiJoinNode()} function, duplicate cases are initially addressed by checking whether the current node has already been processed. If it has, the function returns the result to prevent redundant processing. For nodes that have not been processed, the \texttt{getMultiJoinNode()} function is called recursively to handle its input nodes, and the original input nodes are replaced with the newly returned nodes. When a node is recognized as the root of a multi-way stream node group, the \texttt{createMultiJoinNodeByGroup()} method generates the corresponding multi-way join node, referred to as `multiJoinNode`. The inputs of this new node are aggregated from the member nodes of the group, while excluding the input nodes of the members within the group. This ensures that the multi-way join node accurately encapsulates the connectivity logic of the group and omits nodes that have already been processed. Ultimately, the newly created multi-way stream join node replaces the original root node, thereby forming the new execution plan. This approach guarantees the correct identification and transformation of binary join tree patterns in the execution plan, offering an effective solution for utilizing multi-way stream join operators in stream SQL.

\begin{algorithm}
\caption{getMultiJoinNode}
\label{a3}
\begin{algorithmic}[1]
\Require $node$
\Ensure $ret$
\If{$visitedMap$ contains $node$}
    \State \Return $visitedMap.\text{get}(node)$
\EndIf
\ForAll{$inputNode$ in $node.\text{inputs}$}
    \State $multiJoinNode \gets \text{getMultiJoinNode}(inputNode, visitedMap)$
    \State $node.\text{replaceInputNode}(inputNode, multiJoinNode)$
\EndFor
\State $ret \gets node$
\If{$node.\text{group}$ is not null and $node$ is $node.\text{group}.\text{root}$}
    \State $multiJoinNode \gets \text{createMultiJoinNodeByGroup}(node.\text{group})$
    \State $inputNodes \gets \text{new list of Node}$
    \ForAll{$member$ in $node.\text{group}.\text{members}$}
        \ForAll{$memberInput$ in $member.\text{inputs}$}
            \If{\textbf{not} $node.\text{group}.\text{members}$ contains $memberInput$}
                \State \textbf{add} $memberInput$ \textbf{to} $inputNodes$
            \EndIf
        \EndFor
    \EndFor
    \State $multiJoinNode.\text{setInputNodes}(inputNodes)$
    \State $ret \gets multiJoinNode$
\EndIf
\State $visitedMap.\text{put}(node, ret)$
\State \Return $ret$
\end{algorithmic}
\end{algorithm}

\section{Experiment}
\subsection{Experimental Environment and Data}
A Flink Standalone cluster was established under single-machine configuration, consisting of 8 TaskManagers, with each TaskManager configured with 4 slots. All experiments in this section were conducted with a parallelism level of 32. The hardware and software configuration of the single machine is detailed in Table \ref{tab:hardware_config}.
\begin{table}[htbp]
    \centering
    \caption{Machine Configuration}
    \label{tab:hardware_config}
    \begin{tabular}{ll}
        \hline
        Parameter & Configuration \\
        \hline
        CPU & Intel(R) Xeon(R) Platinum 8163 CPU @ 2.50GHz \\
        Cores & 96 cores \\
        Operating System & Alibaba Group Enterprise Linux Server release 7.2 (Paladin) \\
        Memory & 503GB \\
        Disk & 2.7TB \\
        Java Version & v1.8.0 \\
        Kafka Version & v2.12-3.0.1 \\
        Zookeeper Version & v3.6.3 \\
        Flink Version & v1.17 \\
        \hline
    \end{tabular}
\end{table}
In the experiments of this chapter, the TPC-DS dataset, a standard benchmark in the field of big data decision support, was selected as the reference data. The version of TPC-DS utilized was 2.13.0, which comprises 17 dimension tables and 7 fact tables. Each table averages 18 columns and features multi-dimensional structures, such as snowflake and star schemas, making it highly suitable for online analytical processing. TPC-DS also provides a dataset generation tool (TPC-DS tools) that can create datasets of varying sizes based on different scale factors. For instance, setting the scale factor to 10 generates a dataset of approximately 10 GB. The following four-table join query was executed:
\begin{verbatim}
SELECT *
FROM store_returns sr, customer cu, web_returns wr, catalog_returns cr
WHERE cu.c_current_addr_sk = cr.cr_refunded_addr_sk 
AND cu.c_current_addr_sk = sr.sr_addr_sk 
AND cu.c_current_addr_sk = wr.wr_refunded_addr_sk;
\end{verbatim}
The experiments in this section utilized the TPC-DS dataset with a scale factor of 10. The CSV-formatted data was supplied to the experiment in a random order via Kafka to simulate a real-time data stream. The output results were directed to a black hole to prevent I/O operations from influencing the experimental outcomes. In the TPC-DS dataset with a scale factor of 10, the data information for the four tables involved in the aforementioned query is presented in Table \ref{tab:input_table_info}.
\begin{table}[htbp]
    \centering
    \caption{Input Table Characteristic}
    \label{tab:input_table_info}
    \begin{tabular}{llll}
        \hline
        Table Name & File Size & Data Volume & Tuple Count \\
        \hline
        customer & 64MB & 106MB & 986546 \\
        catalog\_returns & 212MB & 331MB & 2879498 \\
        store\_returns & 323MB & 507MB & 5750864 \\
        web\_returns & 98MB & 149MB & 1438434 \\
        \hline
    \end{tabular}
\end{table}
\subsection{UMJoin Performance}
To evaluate the performance of the UMJoin operator under memory-constrained conditions, a comparison was conducted between the UMJoin operator and the MultiStream operator [42], utilizing an in-memory hash table as the state backend. The memory capacity limit of the state backend was set to 2GB, 3GB, and 4GB, respectively, to observe the variations in the number of output results over time. The experimental results are illustrated in Figures 5a, 5b, and 5c. Under the 2GB and 3GB memory conditions, the MultiStream operator was unable to process all the data and terminated prematurely. Specifically, under the 2GB memory condition, the UMJoin operator produced only about two-thirds of the total output results, while under the 3GB memory condition, the MultiStream operator nearly processed all the data but still failed to handle it completely. In contrast to the MultiStream operator, which is unsuitable for scenarios with limited memory capacity over extended periods, the UMJoin operator successfully generated complete processing results across all three memory capacities. This demonstrates its ability to manage longer data streams more effectively under memory-constrained conditions.

To evaluate the performance of the UMJoin operator under memory resource-constrained conditions, it was compared with a binary join tree (BJT) using the same LSM-Tree backend. To enhance the validity of the comparison, the optimal join order configuration (referred to as Best BJT) was selected for the BJT through traversal. To ensure a fair comparison, both methods were executed under a fixed JVM heap size of 10GB and a parallelism level of 32, while mitigating the potential impacts of data skew on runtime. As illustrated in Figure 5d, under a backend memory configuration of 9GB, both the UMJoin operator and the Best BJT exhibited a trend of decreasing output rates after an initial high rate (0-45 seconds), indicating efficient data processing by both operators during the initial stage with adequate memory resources. However, as the experiment progressed, the output rate of the UMJoin operator decreased more gradually, while that of the Best BJT declined more rapidly. This suggests that the UMJoin operator may demonstrate higher efficiency in data processing compared to the corresponding binary join tree when faced with memory resource constraints. Furthermore, the consistency in the number of final output results, combined with the comprehensive theoretical analysis presented in this paper, supports these findings.

\begin{figure}[H]
	\centering  
	\vspace{-0.35cm} 
	\subfigtopskip=2pt 
	\subfigbottomskip=2pt 
	\subfigcapskip=-5pt 
	\subfigure[2GB Memory]{
		\label{4GB Memory-1}
		\includegraphics[width=0.45\linewidth]{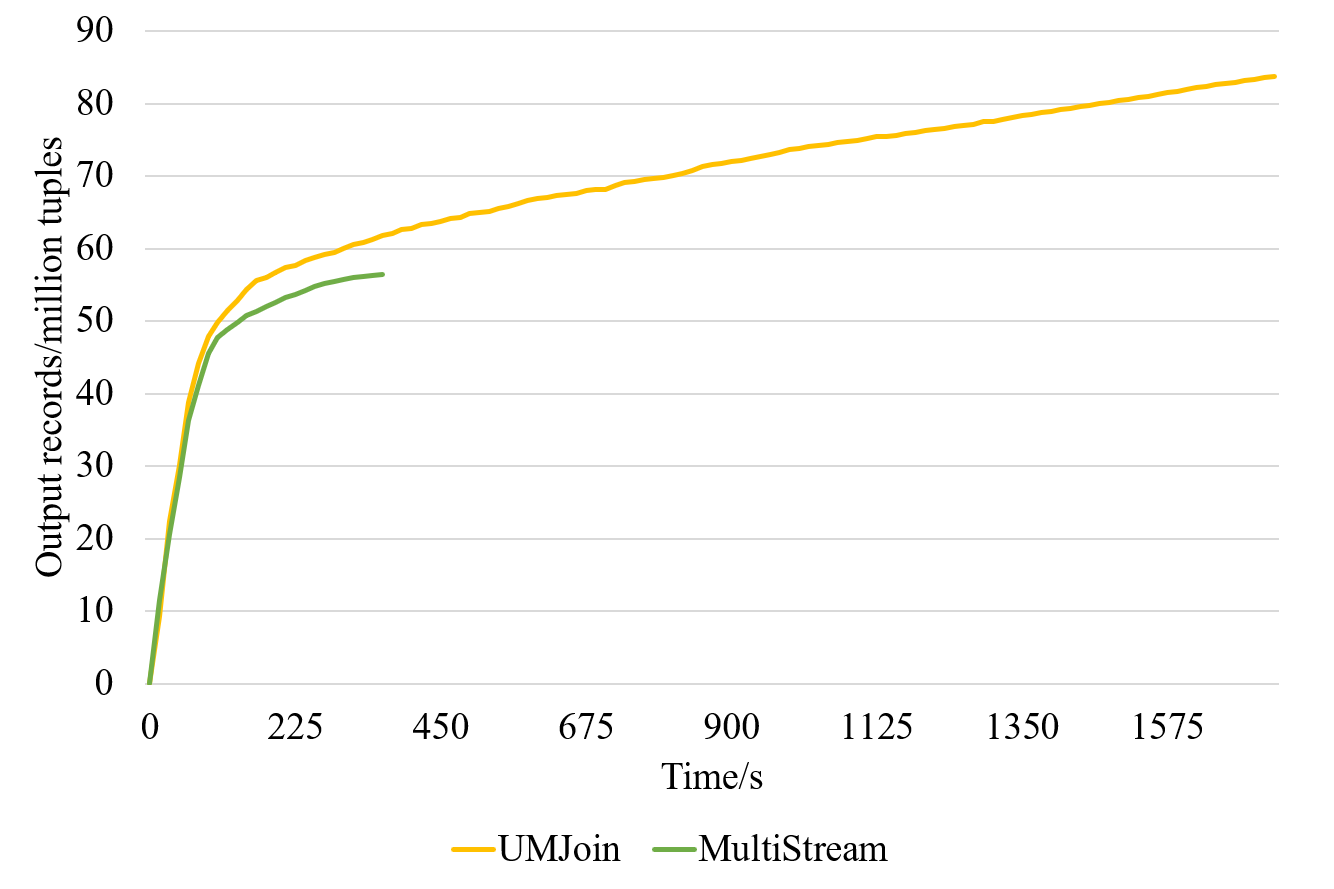}}
	\quad 
	\subfigure[3GB Memory]{
		\label{3GB Memory}
		\includegraphics[width=0.45\linewidth]{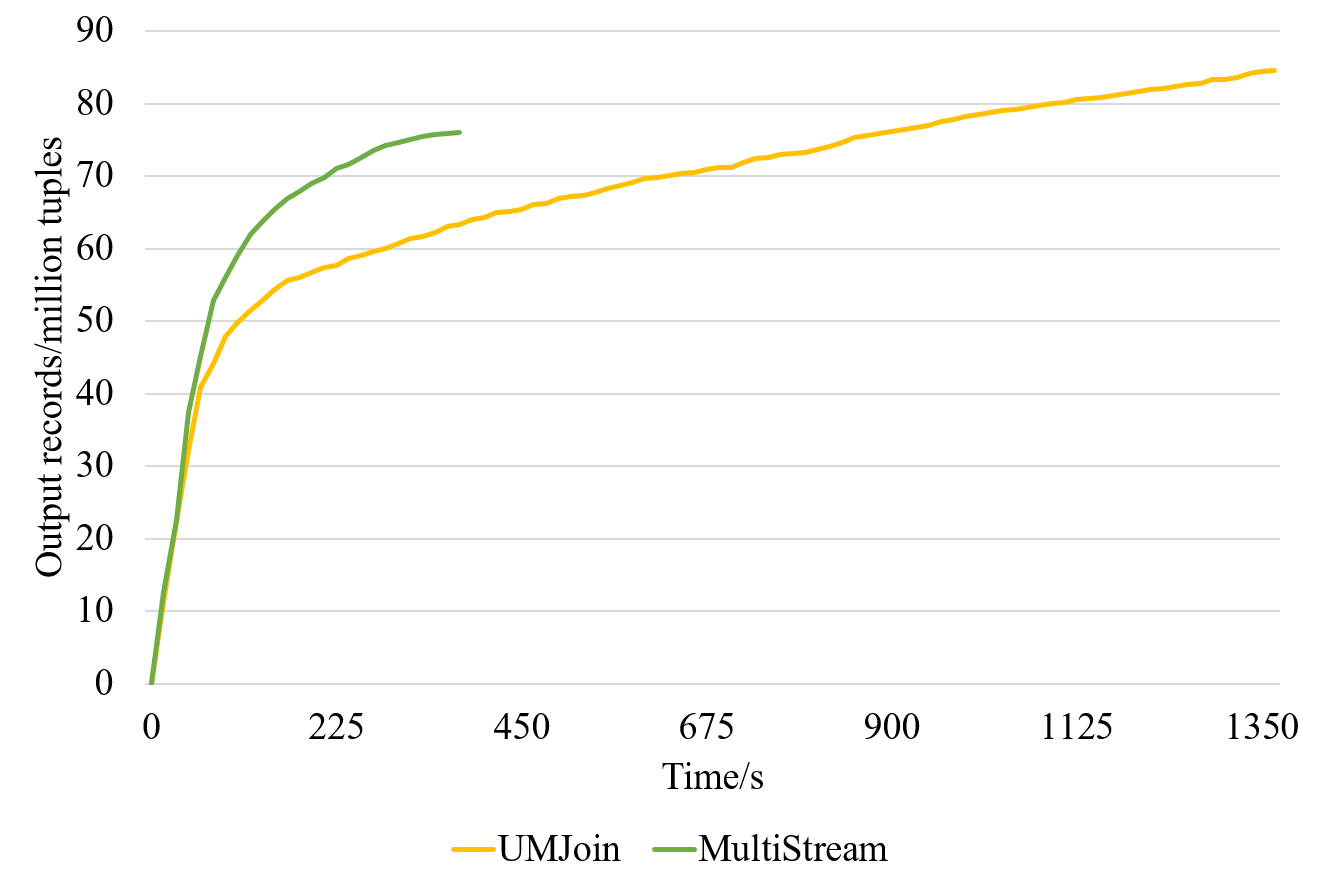}}
	  %这里是空了一行，能够实现强制将四张图分成两行两列显示，而不是放不下图了再换行，使用\\也行。
	\subfigure[4GB Memory]{
		\label{4GB Memory-2}
		\includegraphics[width=0.45\linewidth]{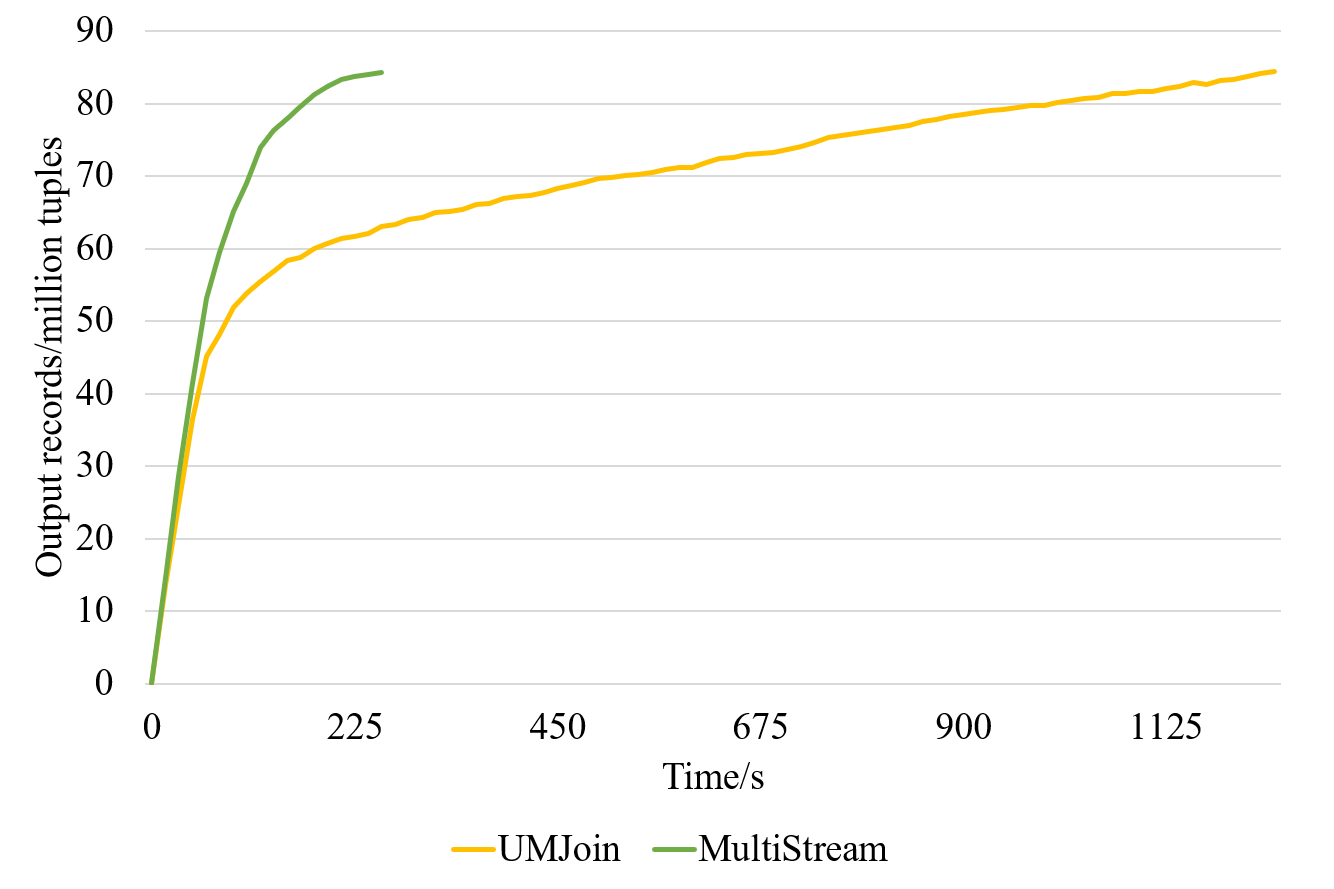}}
	\quad
	\subfigure[UMJoin VS Best BJT]{
		\label{ub}
		\includegraphics[width=0.45\linewidth]{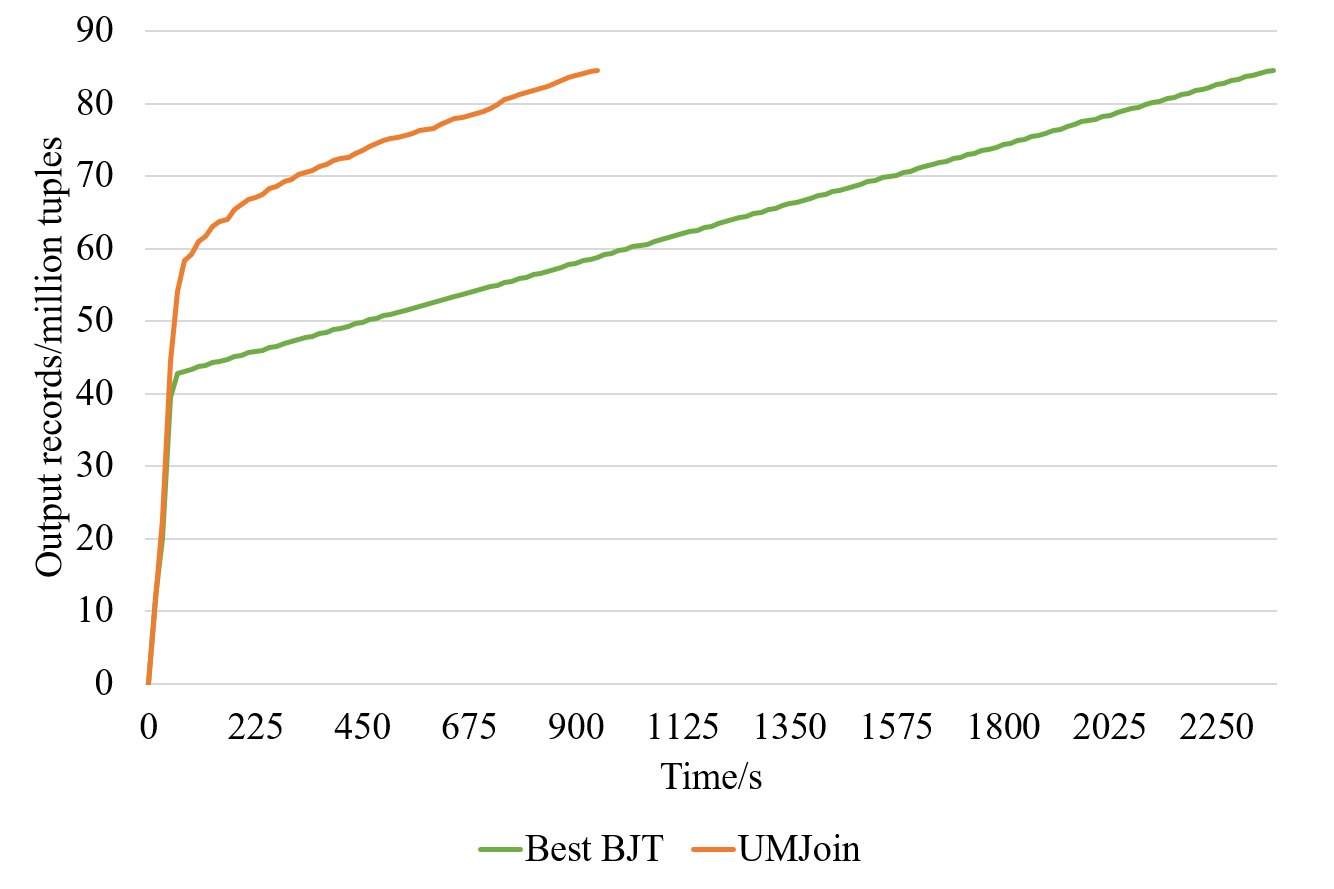}}
	\caption{UMJoin Performance}
	\label{output}
\end{figure}
\begin{figure}[ht]
  \centering
  \includegraphics[width=0.6\linewidth]{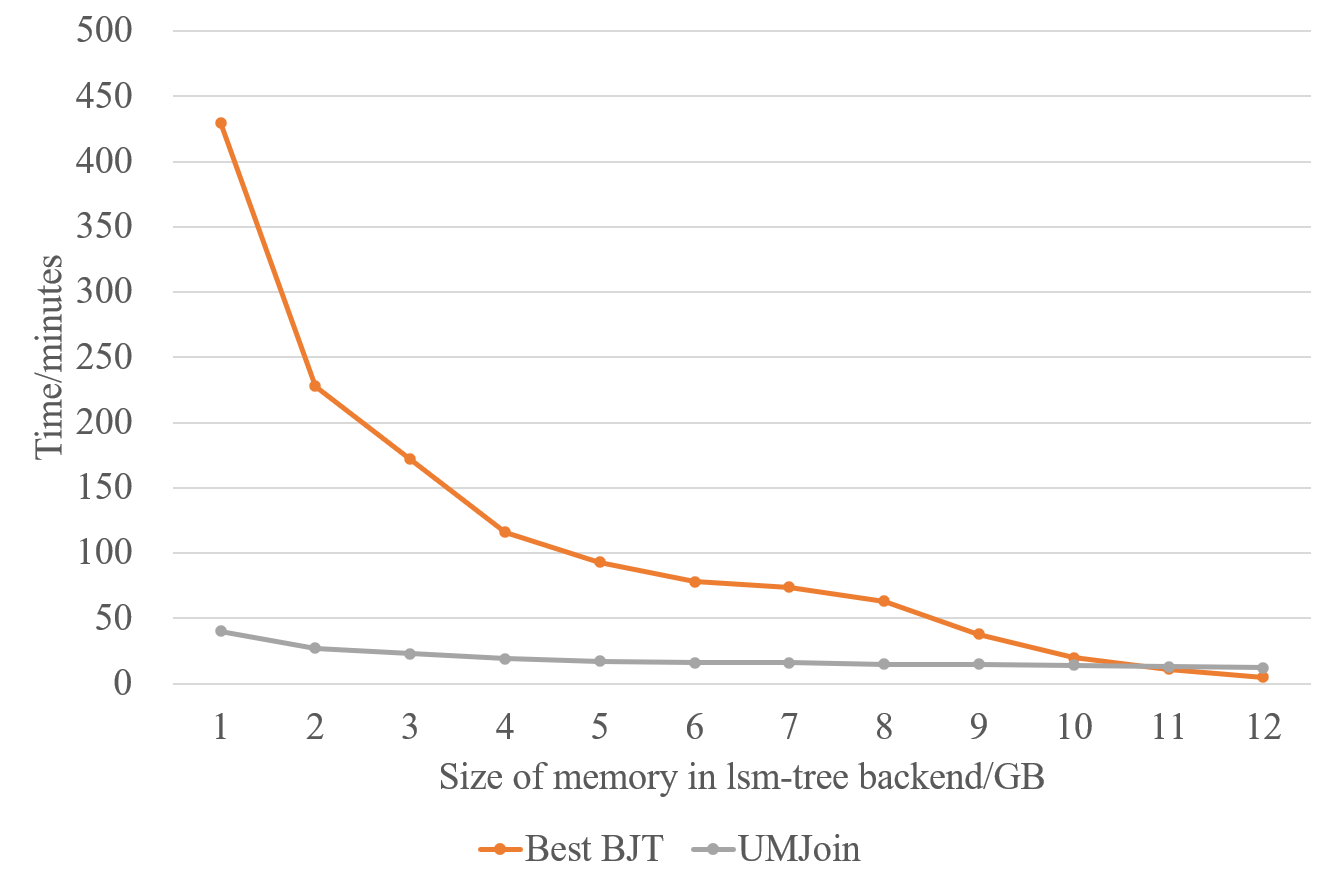}
  \caption{Backend Memory Capacity}
  \label{fig:duration}
\end{figure}
\begin{figure}[H] %这里使用的是强制位置，除非真的放不下，不然就是写在哪里图就放在哪里，不会乱动
	\centering  %图片全局居中
	\vspace{-0.35cm} %设置与上面正文的距离
	\subfigtopskip=2pt %设置子图与上面正文或别的内容的距离
	\subfigbottomskip=2pt %设置第二行子图与第一行子图的距离，即下面的头与上面的脚的距离
	\subfigcapskip=-5pt %设置子图与子标题之间的距离
	\subfigure[Block Cache]{
		\label{Block Cache Capacity}
		\includegraphics[width=0.45\linewidth]{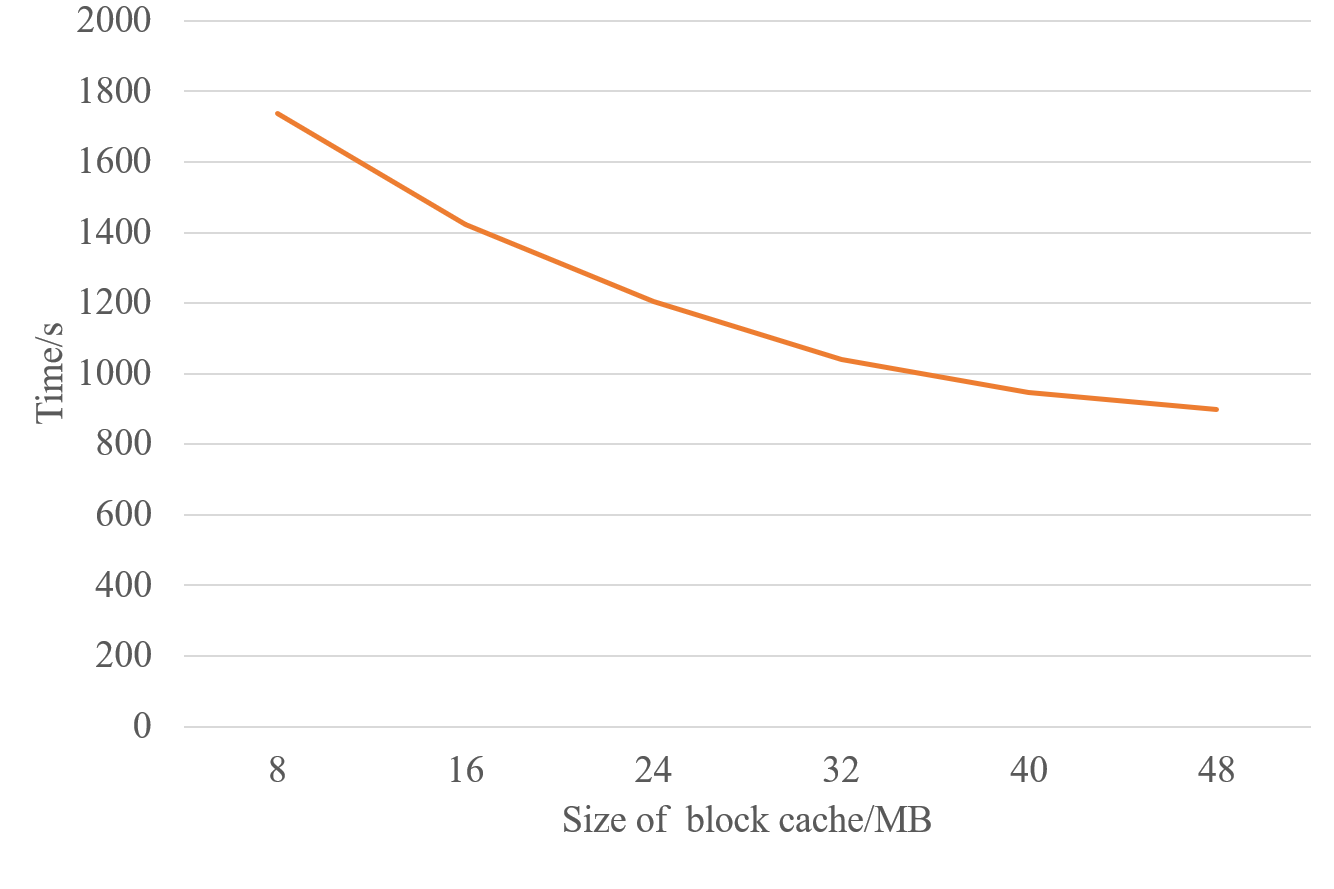}}
	\quad %默认情况下两个子图之间空的较少，使用这个命令加大宽度
	\subfigure[Skiplist Capacity]{
		\label{Skiplist Capacity}
		\includegraphics[width=0.45\linewidth]{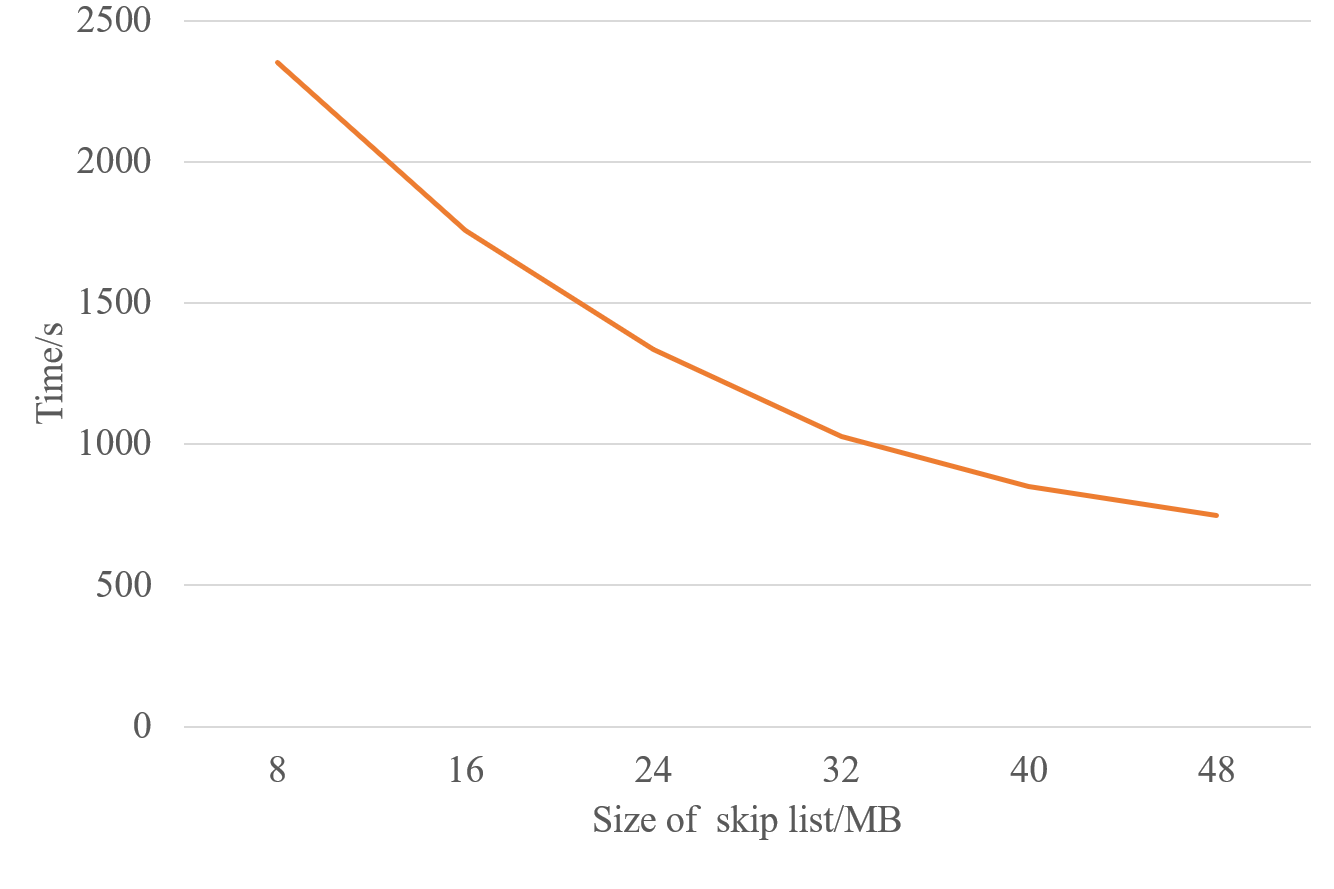}}
	\caption{LSM-Tree Backend}
	\label{lsm}
\end{figure}

To further analyze the impact of memory conditions on the UMJoin operator, we compared the runtimes of UMJoin and Best BJT under various backend memory configurations, ranging from 1GB to 12GB. The experimental results, illustrated in Figure \ref{fig:duration}, indicate that the runtimes of both UMJoin and Best BJT decreased as the backend memory increased. Notably, the runtime trend of the UMJoin operator exhibited a more gradual decline, suggesting its lower sensitivity to changes in backend memory size. This observation underscores the advantage of the UMJoin operator in minimizing the need for intermediate state storage. Under backend memory configurations ranging from 1GB to 10GB, the UMJoin operator consistently outperformed the Best BJT, showcasing its performance advantage in memory-constrained conditions. The Best BJT required an additional 5.03GB of intermediate result data compared to the UMJoin operator. Although the performance of the Best BJT exceeded that of the UMJoin operator when the backend memory configuration increased to 11GB and 12GB, this may be attributed to factors such as the UMJoin operator's failure to cache intermediate results, leading to duplicate calculations, an increased burden on disk read/write operations, and possibly not selecting the optimal detection sequence. However, this does not diminish the value of the UMJoin operator in memory-constrained scenarios. Designed to minimize intermediate result data, the UMJoin operator demonstrates its core advantage in data processing and exhibits superior performance compared to the binary join tree based on the same LSM-Tree backend in memory-constrained environments. This further highlights its potential application in handling multi-stream join queries within streaming processing systems.
Furthermore, to evaluate the impact of changes in block cache and skiplist capacities in the LSM-Tree backend on the efficiency of the UMJoin operator, controlled variable experiments were conducted. The block cache size was fixed at 16 MB while the skiplist capacity was adjusted, and vice versa. The corresponding runtimes are illustrated in Figures \ref{Block Cache Capacity} and \ref{Skiplist Capacity}, respectively. The experimental results indicate that the efficiency of the UMJoin operator is correlated with both block cache and skiplist capacities in the LSM-Tree backend. Specifically, larger block cache and skiplist capacities lead to shorter runtimes. An increase in block cache capacity enhances the amount of read cache data stored in memory, effectively reducing the number of reads from disk SST. Similarly, an increase in skiplist capacity not only increases the cached data in memory but also decreases the frequency of flushing the read-only skiplist to disk SST, thereby reducing write operations to disk. Consequently, when the backend memory of the UMJoin operator is limited, a judicious allocation of skiplist and block cache capacities can significantly improve the operator's runtime efficiency.

\subsection{TSC Method}
After evaluating the performance of the UMJoin operator, further experiments were conducted to validate the effectiveness of the TSC method in transforming execution plans based on binary join trees into execution plans that incorporate integrated multi-stream join nodes. Utilizing the Apache Flink v1.17 framework, the TSC method was implemented through its Flink SQL API [100]. Given the complexity of execution plans and the predicate pushdown optimization capabilities inherent in Flink SQL, the canBeMultiJoinNode() function of the TSC method was enhanced to ensure accurate identification and processing of binary join tree patterns that include projection nodes. In this pattern, the physical execution operator appends a summary field projection operation to the original computation logic, thereby ensuring the validity of the corresponding multi-stream join node. For the selection of benchmark tests, 11 queries containing multi-stream join operations were chosen from the TPC-H benchmark [101,102], which consists of 22 queries for validation. These selected queries include Q2, Q3, Q5, Q7, Q8, Q9, Q10, Q11, Q15, Q18, and Q21, encompassing various forms and complexities of joins.

After inputting the corresponding SQL queries and generating execution plans, the primary evaluation criterion of the experiment was whether the transformed execution plan maintained logical consistency with the original plan and whether the original binary join tree pattern was successfully replaced with new multi-stream join nodes. Figure 8 illustrates the execution plan conversion process of the TSC method for Q3, where Figure 8a depicts the original execution plan of Q3, and Figure 8b shows the new execution plan of Q3 after conversion using the TSC method. It is evident that the TSC method accurately identified the binary join tree pattern in the original execution plan and replaced it with multi-stream join nodes (MultiJoin) without altering the overall execution plan logic, indicating a successful plan conversion.

\begin{figure}[H]
	\centering
    \label{q3}
	%\vspace{-0.35cm}
	% \subfigtopskip=2pt
	% \subfigbottomskip=2pt
	% \subfigcapskip=-5pt
	\subfigure[Q3 Original Execution Plan]{
		\label{Q3 Original Execution Plan}
		\includegraphics[width=0.35\linewidth]{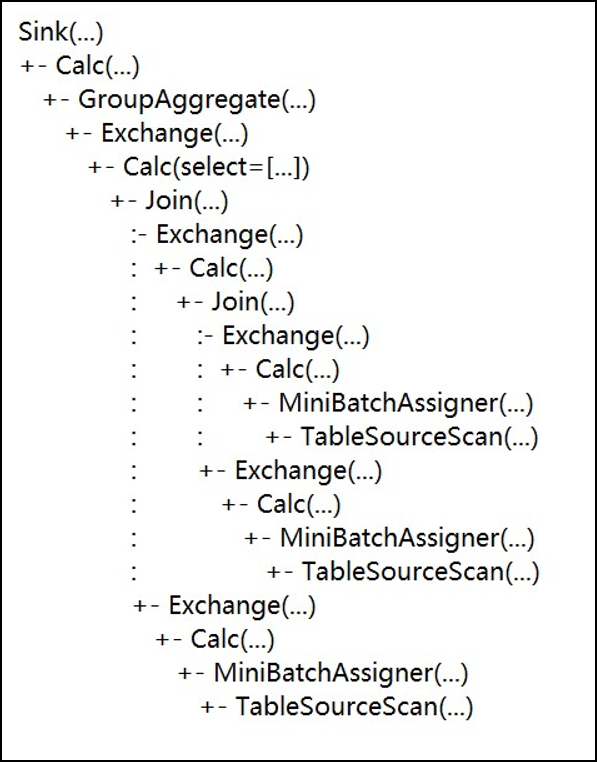}}
	\quad
	\subfigure[Q3 New Execution Plan]{
		\label{Q3 New Execution Plan}
		\includegraphics[width=0.35\linewidth]{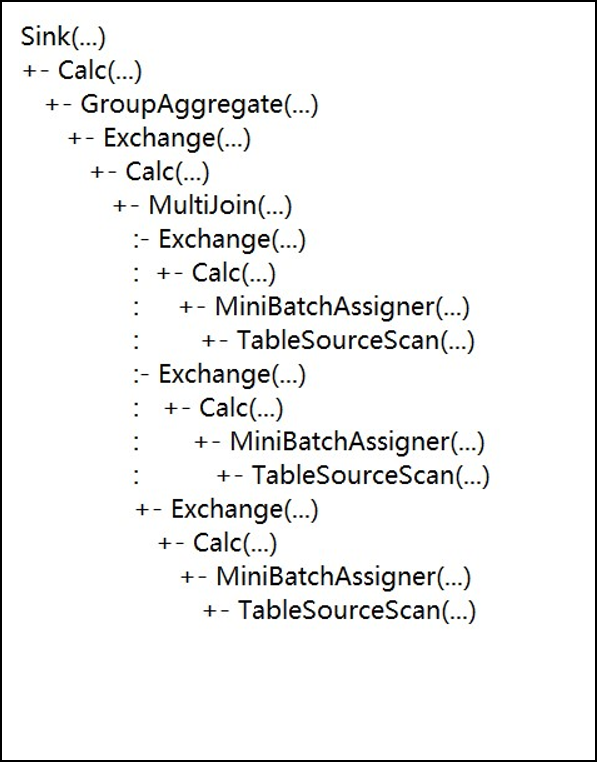}}
	\caption{Conversion of Q3 Execution Plan}
\end{figure}

All conversion results are presented in Table 3, which displays the execution plans of 11 SQL queries that included multi-stream joins. These plans were successfully transformed into execution plans featuring integrated multi-stream join nodes, achieving a pattern recognition and plan conversion accuracy rate of 100\%. This experiment demonstrates the effectiveness and accuracy of the proposed TSC method.

\begin{table}[h!]
    \centering
    \caption{TSC Method Verification}
    \label{TSC_verification}
    \begin{tabular}{c c}
        \toprule
        \textbf{SQL Query Statement} & \textbf{Conversion Result} \\
        \midrule
        Q2  & \checkmark \\
        Q3  & \checkmark \\
        Q5  & \checkmark \\
        Q7  & \checkmark \\
        Q8  & \checkmark \\
        Q9  & \checkmark \\
        Q10 & \checkmark \\
        Q11 & \checkmark \\
        Q15 & \checkmark \\
        Q18 & \checkmark \\
        Q21 & \checkmark \\
        \bottomrule
    \end{tabular}
\end{table}
\section{Conclusion}
This paper addresses the challenges of limited memory capacity and the complexities associated with multi-stream join operators in streaming SQL for processing long data streams by introducing a multi-stream join technique based on the LSM-Tree. We begin by developing a multi-stream join operator (UMJoin) that leverages the LSM-Tree state backend, which utilizes disk storage to enhance the storage capacity for multi-stream states. This approach allows for the use of disk space in conjunction with memory for storing multi-stream state data. Next, we develop a method for executing plan transformation (TSC) for the UMJoin operator, based on the streaming SQL execution process. This method involves identifying binary join tree patterns and creating corresponding multi-stream join nodes. This approach enables the execution plan derived from the binary join tree to be converted into a unified execution plan that incorporates UMJoin nodes. This transformation enhances the application of UMJoin operators in streaming SQL. Experimental results using the TPC-DS dataset demonstrate that the UMJoin operator can effectively and efficiently manage long data streams, even under memory constraints, highlighting its capability to optimize binary join trees with the same LSM-Tree backend. Furthermore, the study investigates how the capacities of skip lists and block caches influence runtime efficiency in the LSM-Tree backend. Concurrently, experiments on the execution plan transformation of multi-stream join queries utilizing the TPC-H benchmark validate the effectiveness of the TSC method in executing these transformations. Future research will focus on addressing data distribution challenges associated with various connection properties in concurrent systems.

\bibliographystyle{unsrt}
\bibliography{references.bib}  %%% Uncomment this line and comment out the ``thebibliography'' section below to use the external .bib file (using bibtex) .





% \end{thebibliography}

\end{document}